\documentclass[a4,10pt,fleqn]{article}
\usepackage{amsmath,amsfonts}
\usepackage{algorithmic}
\usepackage{algorithm}
\usepackage{array}
\usepackage[caption=false,font=normalsize,labelfont=sf,textfont=sf]{subfig}
\usepackage{textcomp}
\usepackage{stfloats}
\usepackage{url}
\usepackage{verbatim}
\usepackage{graphicx}
\usepackage{color}
\usepackage{cite}
\usepackage{comment}
\usepackage{siunitx}
\usepackage{amsmath,amssymb, array, cases,mathrsfs, here,enumitem,ascmac,url, cite, float}
\usepackage{multirow}
\usepackage[margin=20truemm]{geometry}

\newcommand{\fref}[1]{Fig.~\ref{#1}}

\newcommand{\tref}[1]{Table~\ref{#1}}

\begin{document}

\title{Router for Wireless Power Packet Transmission: \\Design and Application to Intersystem Power Management\thanks{This work has been submitted to the IEEE for possible publication. Copyright may be transferred without notice, after which this version may no longer be accessible.}}

\author{Takahiro Mamiya, Shiu Mochiyama, and Takashi Hikihara\\\\
Department of Electrical Engineering, Kyoto University}

\date{}

\maketitle

\begin{abstract}
Power supply for small-scale battery-powered systems such as electric vehicles (EVs and mobile robots) is being actively researched.
We are particularly interested in energy management, which considers the interconnection of such systems close to each other.
This allows for overall redundancy to be maintained without assuming excessive redundancy with individual power sources.
Its implementation necessitates a high level of integration between power management and information and communication technology. 
As one of these methods, this study investigates energy management based on power packetization. 
When the individual systems to be connected have moving parts or are mobile, wireless power transmission is a promising method for power sharing.
However, power packetization has so far only been considered for wired transmission.
In this paper, we address the integration of power and information in wireless channels using power packetization.
We propose a power packet router circuit that can wirelessly transmit power over multiple channels selectively.
Furthermore, we demonstrate that the developed system can handle both wired intrasystem power management and wireless intersystem power sharing in a unified manner.
\end{abstract}

\section{Introduction}
\label{chap:intro}
Recent days have witnessed widespread use of electric power systems that are equipped with batteries and can thus be driven without relying on an external and large power grid.
 Common examples include electric vehicles (EVs) and mobile robots.
While much effort has been dedicated to independent power management in such a system, another research trend is the management of a network of such systems.
We refer to a minimum element of a system that can independently operate a local system throughout the paper.
Constituting a networked system addresses shared redundancy of power source capacity as a whole system, rather than as each individual system.
That is, when the power demand of one system temporarily increases, 
power can be supplied not only from the inside power sources but also from the power sources of the other connected systems\cite{Dialynas2007, He2008, Farhadi2015}. 

Because local systems are spatially dispersed and can have a time-dependent supply/load profile, managing such a network necessitates advanced sensing, computation, and communication technologies\cite{Wu2020, smartgrid, Extended}.
Several proposals for power system management with ICTs support have been made\cite{He2008, Sugiyama2017, Gelenbe2016}.
Among them, a power packet dispatching system is an encouraging proposal for the purpose. 
The system packetizes supplied power; that is, power is divided into time segments, each of which is associated with an information tag via a voltage waveform\cite{Takahashi2015a, Takahashi2017}.
Power packetization ensures that information exchange and power transmission occur concurrently in the physical layer, allowing for power management in a network without causing an imbalance in information and physical quantity processing. 
In the previous study, the authors' group developed a circuit called a power packet router\cite{Takahashi2015a}.
We validated the concept of power packetization and routing with hardware configuration including the routers.

One advantage of power packetization is the ability to easily attach/detach local systems from a larger network.
The use of time-division multiplexing and physical tag attachment ensures that each packetized power transfer is independent.
In other words, power transfers between different pairs do not get mixed up even on the same power line but can be differentiated physically.
This leads to realizing what could be called a plug-and-play from the perspective of power supply.

One difficulty here is that the power packet dispatching system has so far been developed using a wired connection for power transfer.
Wireless power transfer (WPT) is a revolutionary technology for supplying power to mobile systems\cite{WiTricity, Karalis2008}.
It is beneficial for improved maneuverability of each local system to introduce the WPT to the power packet dispatching system for connections at the boundaries of the local systems.
However, as discussed in Section~\ref{Router design for wireless transmission of power packets}, simply connecting a WPT circuit to a power packet system does not work in conjunction with packet-based power management in local systems.
This research seeks to achieve on-demand power supply concentration and dispersion in the connected network of local systems while ensuring easy attachment/detachment between systems via a wireless connection.

Here, we investigate the following two points as fundamental studies to realize the wireless connection of multiple local systems powered by power packets. 
First, we suggest a dedicated router design in both software and hardware configurations to ensure physical packetization with a wireless channel and collaboration with wired power management. 
Then, in a connected system comprised of three local systems with the developed router installed, we assert the selective transmission of power packets to only the local system designated by the tag. 
Second, we demonstrate a power-sharing strategy for increasing power capacity redundancy via a wireless connection. 
With wired and wireless connections, the parallel operation of intra- and intersystem power management is demonstrated.
Among a network of two local systems, each of which supplies a certain demand of its own load via wired connection, surplus power at one system is transferred to another. 

Many proposals for the duplex of multiple channels in WPT have been made, including multiplexing in time, frequency, and spatial domain\cite{Hou2022, Ota2021, Shinohara2021}. 
Furthermore, several reports have addressed the simultaneous wireless transmission of power and information\cite{WTPI_ASK, Mase2021, Hossain2019, Wu2020}. 
These proposals essentially assume that a power transmission channel has already been established and that information is being transmitted concurrently, or vice versa. 
Our proposal, on the other hand, attempts to go beyond the simple parallel transmission by integrating information that manipulates the spatiotemporal distribution of power with power transmission itself at the physical layer. 
This, in theory, eliminates the disparity between physical quantity and information, allowing us to achieve both wired and wireless power transmission by cooperating for smart power management.

\begin{figure}[t]
  \centering
  \includegraphics[width = 0.6\hsize]{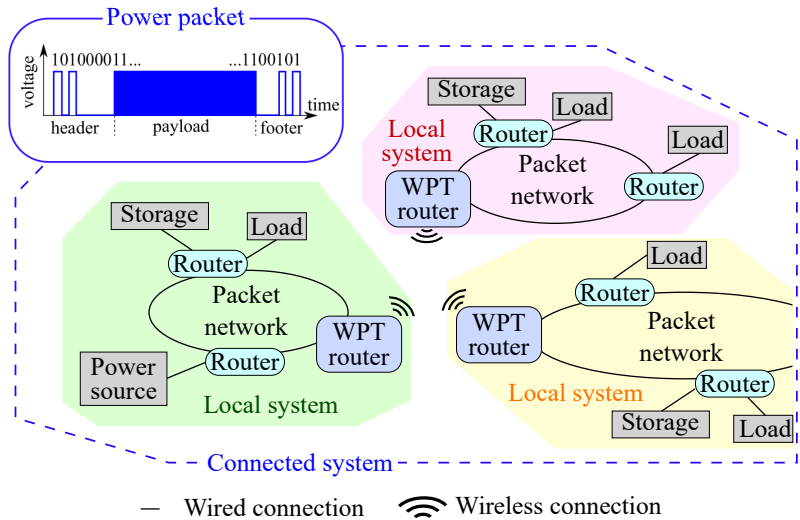}
  \caption{Surplus power supply via wireless power transfer between power packet networks.}
  \label{fig0}
\end{figure}

\section{Outline of power packet dispatching system}
\label{Outline of power packet dispatching system}
The basic configuration and operation of power packet dispatching systems are described in this section.

\subsection{Constitution of power packet}
As depicted in \fref{fig0}, a power packet is a unit of power management in the system. 
A power packet comprises pulse-shaped electric power called a payload and an information tag, a header, and a footer, which are attached just before and after it.
The information tag is a logic bitstream realized by a voltage waveform without current.
The tag can include any information, like the origin, destination, and length of the power packet.

The physical tag attachment enables power packet transmission to be time-division multiplexed.
Power from different sources and destinations is transmitted on the same channel while remaining distinct from one another.
This feature sets the power packet dispatching system apart from conventional systems that treat power as a continuous flow.

\subsection{Network configuration of power packet dispatching system}
\label{Network configuration of power packet dispatching system}
Each local system of \fref{fig0} denotes a local system configuration example.
Routers connect power sources, storage, and loads to the network in this system.
A power packet router is installed as a node that connects multiple transmission lines.
The router forwards power packets by selecting a transmission line according to the packet's tag information \cite{Takahashi2017}.

A power packet is routed from a source to a specific destination via several routers.
The path to the load is not required to be unique and can be changed dynamically depending on the situation. 
This feature facilitates flexible power management in conjunction with a dynamic supply relationship. 
In the following section, we develop a router that can perform this function even with a wireless connection.

\subsection{Routing method for power packets}
\label{Routing method for power packets}
\begin{figure}[t]
  \centering
  \includegraphics[width = 0.5\hsize]{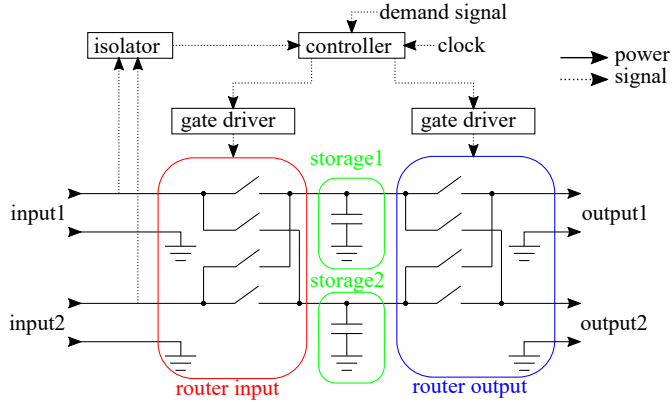}
  \caption{Circuit example of 2 input 2 output router\cite{Takahashi2015a}.}
  \label{fig1}
\end{figure}
Here we characterize the circuit configuration of a router and the principle of  
its routing operation\cite{Takahashi2015a,Inagaki2020,Mochiyama2019}.
Figure~\ref{fig1} depicts the circuit configuration of a previously proposed, wire-connected router\cite{router}.
The circuit consists of two sections: an input section that receives power packets from the transmission network and an output section that forwards power packets to the transmission network.
The operation of the input part is initialized when a power packet reaches the router.
The input section includes a signal reading circuit for reading the logic signals of information tags.
When the router recognizes that the incoming power packet is addressed to it, it turns on the corresponding semiconductor switches to receive the payload power.
For circuit protection, the signal reading circuit electrically separates its signal output from the power supply lines using a device such as a photocoupler.
The incoming power packet is temporarily stored before being forwarded to the next hop. 
The output part generates power packets from the temporal storage in response to the demand.
In some cases, the circuit can be reduced to just the input or output section. 
When installed just next to the source, for example, the output section with the storage replaced by a power source is sufficient to produce a generated packet. 
Similarly, a circuit just before a load can only be the input part, with the storage replaced by a load. 

To read the logic signals of power packets, clocks corresponding to the one-bit width of a power packet must be synchronized among adjacent routers.
This can be accomplished by installing an additional wire for a common clock input, or by adding another signal to the header for autonomous clock synchronization \cite{zhou-clock}.
In this paper, we employ a simple autonomous clock synchronization scheme, in which the clock period is fixed in advance, the first three bits of the header are set to 010, and the phase is shifted if the 010 is not detected within a certain period.

The information tag consists of bits 1--3, which implies 010 for clock synchronization, and bits 4--7, which imply the address of the output destination.
Bits 8 -- 100 correspond to the payload.
For simplicity, the packet length is fixed at 100 bits and this setting is shared by all routers.
In this way, we exclude the footer.

\section{Router design for wireless transmission of power packets}
\label{Router design for wireless transmission of power packets}
In this section, we propose a router configuration for wireless transmission of power packets. 
We employ magnetic resonant coupling for the wireless transmission.
This method is capable of transmitting large power over long distances with high-efficiency \cite{WiTricity}.
\begin{figure}[b]
  \centering
  \includegraphics[width = 0.5\hsize]{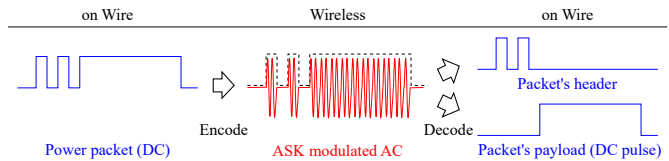}
  \caption{A waveform concept during wireless transmission of power packets.}
  \label{fig2}
\end{figure}
This circuit is powered by AC, whereas the power packet dispatching system is powered by DC.
We must convert the current to incorporate WPT into power packet routing. 
Figure~\ref{fig2} depicts a conceptual diagram of the voltage and magnetic flux density in the wireless power packet transmission. 
Using a magnetic resonant coupling circuit that includes an inverter and a rectifier, DC is converted to AC and then back to DC after wireless transmission.
The following section describes the router design. 

It should be noted that the inclusion of wireless transmission in the power packet dispatching system was first proposed in the authors' previous report \cite{GCCE}. 
In the report, the wireless transmission was not packetized but introduced as a one-to-one transmission channel without any tag attachment. 
In this paper, we propose a novel router configuration that bring the functions of physical tag attachment and its reading to the wireless power transfer. 
These functions not only realize the physical packetization of wireless power transmission but also extends its use to packet-based power management as introduced in Section~\ref{Confirmation of power-sharing in the wirelessly connected systems}. 

\subsection{Wireless transmitter of the power packet}
\label{Wireless transmitter of the power packet}
\begin{figure}[t]
  \centering
  \includegraphics[width = 0.3\hsize]{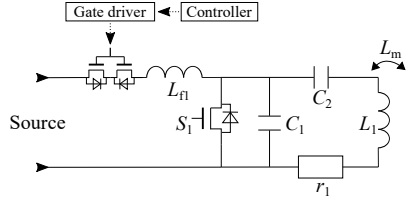}
  \caption{Wireless transmitter of the power packet.}
  \label{fig3}
\end{figure}
Figure~\ref{fig3} depicts a router circuit for wireless power packet output.
The configuration includes an inverter circuit connected to the router's output section, as described in Section~\ref{Routing method for power packets}.
For DC/AC conversion, a class-E inverter \cite{RPC} is used.

The output circuit wirelessly transmits both the header signal and the payload power. 
In this paper, the inverter's input is presented as a form of packetized power.
The current flowing through the coil and the magnetic flux density induced in the coil is thus modulated in an amplitude-shift-keying (ASK) manner according to the shape of the power packet, as depicted in the middle of \fref{fig2}.
It should be noted here that the header signal transmission must minimize power consumption while the payload transmission must maximize the amount of power transferred. 
The two requirements cannot be met solely through the transmitter's operation, but rather through the design of the receiver side. 
This point will be covered in greater detail in the following section.

\subsection{Wireless receiver of the power packet}
\label{Wireless receiver of the power packet}
\begin{figure}[t]
  \centering
  \includegraphics[width = 0.3\hsize]{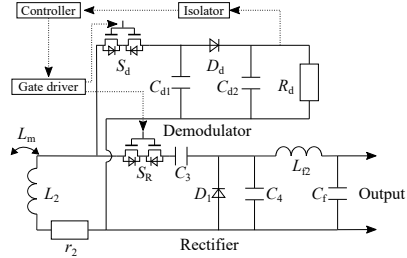}
  \caption{Wireless receiver of the power packet.}
  \label{fig4}
\end{figure}
To receive a wirelessly transmitted power packet, demodulation of the ASK-modulated header signal and highly efficient AC/DC conversion of the payload are necessitated.
The proposed circuit shown in \fref{fig4} meets both requirements by dividing the demodulator into two circuits.
The signal demodulation circuit reads the header, and a class-E rectifier receives the payload.
The detailed procedure is provided below.
Initially, the switch $\mathrm{S_\mathrm{d}}$ connected to the signal demodulation circuit is turned on, while the $\mathrm{S_\mathrm{R}}$ connected to the rectifier circuit is turned off.
For signal demodulation, the envelope of the voltage across the secondary circuit's resonant capacitor is passed through an RC low-pass filter. 
The router's controller then samples it at a predetermined clock cycle to convert it into a logical sequence. 
The controller activates the switches that connect the coil to the rectifier circuit when it determines from the tag that the power packet is addressed to itself.
This causes a class-E rectifier to convert the wirelessly transmitted payload into DC output.
When the power packet is directed at another router, the router's controller disconnects both circuits and opens the coil. 
The detachment is used to avoid the influence of the unintended connection and the resulting impedance change, which may degrade power transmission at the addressed connection. 
At the end of the previous power packet, the controller turns on the switch to the demodulation circuit to prepare for the next power packet. 
The end of a power packet is detected by simply counting the length of the payload in 100-bit intervals.

Of course, simply connecting the signal demodulation circuit and the Class-E rectifier in parallel allows you to read the header and receive the payload. 
However, when receiving the header, the current passes through the Class-E rectifier, and when receiving the payload, it passes through the demodulation circuit. 
Such a current contributes nothing to the receiving operation but results in power loss. 
Because this type of loss is much greater than the loss caused by the switching of the two demodulation circuits, the proposed scheme can greatly reduce the loss. 

The frequency of the carrier wave used for magnetic resonant coupling is \SI{1}{MHz}.
The wireless router's constants are determined as shown in \tref{character}.
The design is conducted in the following manner, regarding \cite{RPC}. 
The coil has a diameter of \SI{100}{mm}, a wire diameter of \SI{1}{mm}, several turns of 10, and a thickness of \SI{12}{mm}. 
The transmission circuit's rise time was measured to be \SI{25}{\mu s}.
The rise time is defined as the time required for the output voltage to attain \SI{90}{\%} of its steady-state value. 
The steady-state value was obtained under the test condition where the load was \SI{47}{\ohm} resistor and the vertical distance between the coils was \SI{50}{mm}.
Based on this, we determined that the bit width of the power packet should be \SI{100}{\mu s}, which is sufficiently larger than the rise time.
That is, the modulation frequency is \SI{10}{kHz}.
The demodulation circuit is designed to demodulate signals with a cutoff frequency of about \SI{100}{kHz}.

\begin{table}[h]
  \caption{Design values of circuit constants.}
  \label{character}
    \begin{center}
      \begin{tabular}{cc|cc|cc}
        \hline
        \multicolumn{2}{c|}{\multirow{2}{*}{Primary side}}    &\multicolumn{4}{c}{Secondary side}                 \\ \cline{3-6}
        &&\multicolumn{2}{c|}{Rectifier} & \multicolumn{2}{c}{Demodulator}                                       \\ \hline

        $f$             & \SI{1}{MHz}          &$L_2$                          & \SI{19.2}{\mu H}                &$C_\mathrm{d1}$ & \SI{1.0}{\mu F}            \\ 
        $L_\mathrm{f1}$ & \SI{100}{\mu F}      &$r_2$                          & \SI{0.88}{\ohm}                 & $C_\mathrm{d2}$ & \SI{820}{\pico F}            \\ 
        $C_1$           & \SI{3.3}{\nano F}    &$C_3$                          & \SI{1.56}{\nano F}              & $R_\mathrm{d}$  & \SI{12}{k \ohm}            \\ 
        $C_2$           & \SI{1.44}{\nano F}   &$C_4$                          & \SI{1.68}{\nano F}              & &            \\ 
        $L_1$           & \SI{19.3}{\mu H}     &$L_\mathrm{f2}$                & \SI{100}{\mu H}                    &&                 \\ 
        $r_1$           & \SI{0.88}{\ohm}      &$C_\mathrm{f}$                 & \SI{0.47}{\mu F}                & &            \\ 
        $L_\mathrm{m}$  & \SI{1.75}{\mu H}     &&&            \\ \hline\\
      \end{tabular}
    \end{center}
\end{table}

\section{Verification of selective reception of wirelessly transmitted power packets}
\label{Verification of selective reception of wirelessly transmitted power packets}
In this study, we consider one-to-many or many-to-many wireless power sharing among several local systems placed close to each other. 
The packetization and time-division multiplexing methods enable simultaneous supplies between different pairs of a transmitter and a receiver while completely distinguishing them. 
Here, we experiment with three local systems, one transmitting and two receiving nodes. 
It is demonstrated that the two receivers can selectively accept or reject power packets based on the attached information tag. 
The number of local systems and their connection relationship can of course be easily expanded and modified due to packetization.

\subsection{Experimental setup for selective reception}
\label{Experimental setup for selective reception}
\begin{figure}[t]
  \centering
  \includegraphics[width = 0.8\hsize]{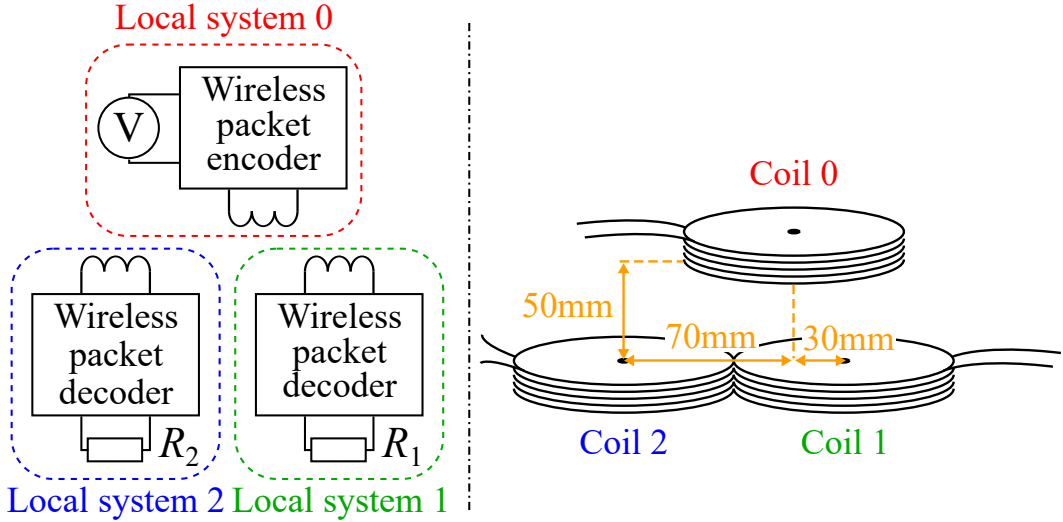}
  \caption{Network configuration with 3 local systems for verification of selectivity of wirelessly transmitted power packet.}
  \label{fig5}
\end{figure}
The entire network configuration is depicted in \fref{fig5}.
Local system 0 alternately sends power packets to local systems 1 and 2, and systems 1 and 2 receive only those that match their addresses. 
Power packet header addresses are set to 0001 and 0010 for systems 1 and 2, respectively.
Local system 0 consists of a circuit from \fref{fig3} and a DC power supply of \SI{12}{V}. 
Local systems 1 and 2 comprise a circuit of \fref{fig3} with a load resistor of \SI{47}{\ohm} connected to the output port. 

Although the transmitting and receiving roles of the local systems are fixed for simplicity, it is possible to transmit power packets bidirectionally by modifying the circuit configuration\cite{Hosotani2012}.
Therefore, this assumption will not lose generality in power sharing. 

To ensure that the router's operation is not affected by the distance between the coils, the coil positions are set as shown in \fref{fig5}.
The coils of local systems 1 and 2 are placed at the same vertical distance \SI{50}{mm} as the coil of local system 0, but the horizontal distance is \SI{30}{mm} and \SI{70}{mm}, respectively.

\subsection{Receiving mode confirmation}
\label{Receiving mode confirmation}
First, we examine the switching behavior between the header signal demodulator and the payload rectifier circuit, as designed in Section~\ref{Router design for wireless transmission of power packets}. 
\begin{figure}[t]
  \centering
  \includegraphics[width = 0.6\hsize]{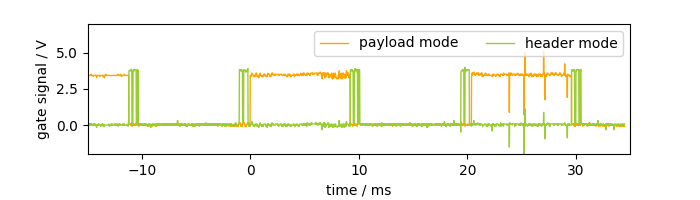}
  \caption{Verification of router operation mode for wirelessly transmitted power packets.}
  \label{fig6}
\end{figure}
Figure~\ref{fig6} depicts an internal signal of the router of local system 1 that represents the receiver's operation mode.
The router was in the header mode every \SI{10}{ms}, which corresponded to the transmission cycle of the powder packets.
Immediately after the header mode, the router switched to the payload mode every two power packet deliveries.
During the payload mode, power was supplied to the designated load. 
This suggests that the controller received the header while connected to the demodulation circuit and then switched to the rectifier circuit in payload mode after recognizing the address.
This result confirms that the proposed router can correctly route power packets on the wireless channel.

\subsection{Confirmation of selective reception function}
\label{Confirmation of selective reception function}
\begin{figure}[t]
  \centering
  \includegraphics[width = 0.6\hsize]{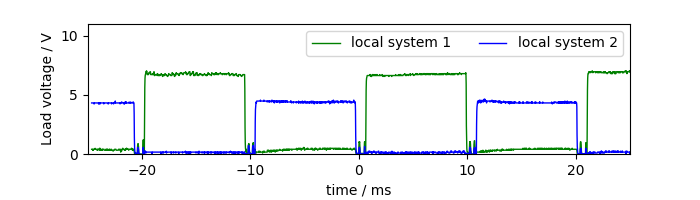}
  \caption{Voltages at two loads in local systems 1 and 2.}
  \label{fig7}
\end{figure}
Second, we confirm that, according to the tag information, the local systems received time-division multiplexed power packets.
The load voltages of the routers of local systems 1 and 2 are depicted in \fref{fig7}.
It can be seen that local systems 1 and 2 received power alternately, indicating that they selectively accepted or denied receiving power packets based on the attached destination address signal. 
Here, local system 1's supply voltage was higher than that of local system 2.
This is because the output is proportional to the distance between the coils. 
This means that, regardless of whether the output value is larger or smaller, the router's selective reception is unaffected by the difference in distance between the coils.

\section{Confirmation of power-sharing in the wirelessly connected systems}
\label{Confirmation of power-sharing in the wirelessly connected systems}
\begin{figure*}[t]
  \centering
  \includegraphics[width = \hsize]{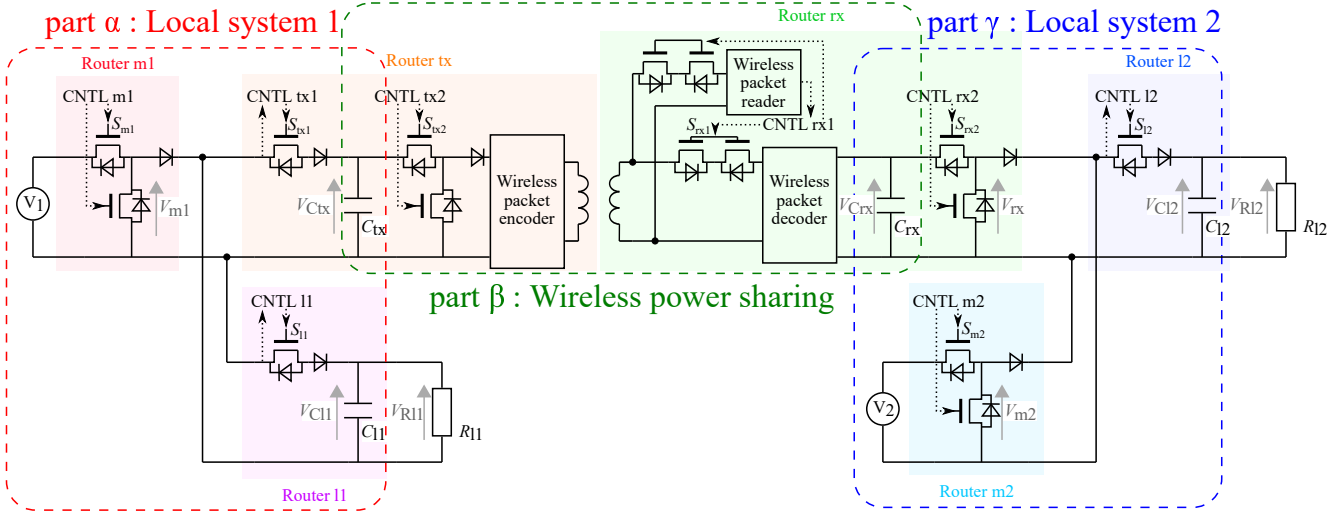}
  \caption{Configuration of the network with 2 local systems connected wirelessly.}
  \label{fig8}
\end{figure*}
Next, in wirelessly connected local systems, we validate power management based on power packetization. 
We consider two local systems where the local power supply is primarily managed via a wired connection. 
Every local system consists of an internal power source, a capacitor, a wireless transmission circuit, and a resistive load.
We set a wirelessly connected networking system comprising two such local systems, as shown in \fref{fig8}.
While each local system supplies its source to its load, wireless power packet transmission compensates for excess or deficient power. 
Each system's goal is to keep the voltage supplied to the load above a certain level.

The proposed scheme deals with a connected system whose elements are subject to dynamic changes, such as variable distance between local systems and time-dependent connection/disconnection of local systems. 
Dealing with such dynamic changes altogether in a centralized controller is not ideal. 
Distributed control of power packet transmission, however, is an effective method of accommodating such unpredictability. 
In this paper, we use a distributed control scheme of packet-based power management\cite{Katayama2018}, in which power packet transmission is managed only between adjacent routers. 
The following section describes the operation flow of the connected systems.

\subsection{Operation flow in connected systems}
\label{Operation flow in connected systems}
Capacitors are installed in the connected systems to generate and output power packets to the load. 
Power packets are sent so that the voltages of these capacitors exceed a certain threshold.

The demand signal to the router for on-demand packet transmission can be given by information tags in power packets or by using another channel such as radio signals\cite{Katayama2018}.
In this paper, we use an external wire to transmit demand signals for simplicity 
We designed an input high to the controller of the next router when the storage voltage falls below the threshold.

We divide the configuration of \fref{fig8} into the following three parts that are managed independently. 
\begin{itemize}
  \item[$\alpha$] Transmission from router m1 to router tx and router l1
  \item[$\beta$] Transmission from router tx to router rx
  \item[$\gamma$] Transmission from router rx and router m2 to router l2
\end{itemize}
The three parts' basic operation principles are described below. 

In part $\alpha$, when the voltages across $C_\mathrm{tx}$ and $C_\mathrm{l1}$ fall below the threshold, demand signals are transmitted to the router m1 respectively. 
Router m1 generates and sends power packets to the destination from which the demand signal is received.
In the event of overlapping demand signals, priority is given to router l1 to keep the load voltage stable. 

In part $\beta$, router rx sends a demand signal to router tx when the voltage across $C_\mathrm{rx}$ falls below the threshold. 
Router tx generates and sends power packets to router rx based on the demand signal. 

In part $\gamma$, when the voltage across $C_\mathrm{l2}$ drops below the threshold, a demand signal is initially sent to router rx. 
If a power packet is not delivered from router rx to router l2 within a certain amount of time, the demand signal is sent to router m2, which generates and sends a power packet to router l2. 

Besides the three principles, two constraints are imposed on the operation of routers tx and rx. 
First, they do not output power packets if the voltages across its capacitor, $C_\mathrm{tx}$ or $C_\mathrm{rx}$, are lower than a certain value.
To transmit power packets, there must be an adequate potential difference between the source and the destination. 
This constraint guarantees the possible difference between the source and destination capacitors and guarantees the reliable transmission of power packets. 
Second, the routers are not enabled to input and output power packets simultaneously. 
When both switches are switched on simultaneously, the circuits before and after the router are linked parallel. 
In this case, the output impedance measured from the power supply (capacitor) located before the router is lower than when only the input switch is turned on. 
This can result in an overcurrent at the source and a rapid drop in capacitor voltage. 
The second constraint is levied to avoid this situation. 
This configuration may prevent the router rx from emitting power on occasion. 
Even if this occurs, router m2 can supply power packets to keep router l2's voltage stable.

\subsection{Verification of connected systems operation}
\label{Verification of connected systems operation}
To test the operation of the connected systems, we set the supply voltages $V_1=$\SI{15}{V} and $V_2=$\SI{7}{V}. 
To create a voltage gradient, the threshold voltages of capacitors $C_\mathrm{l1}$, $C_\mathrm{tx}$, $C_\mathrm{rx}$ and $C_\mathrm{l2}$ are set as \SI{10}{V}, \SI{9}{V}, \SI{7}{V} and \SI{5}{V}, respectively. 
The parameters linked to wireless power transmission are set as depicted in \tref{character} 

It is worth noting that the routers' wired channel switch units have been replaced with unidirectional ones. 
As previously discussed, the symmetry of the circuit allows us to restrict the flow of power packets to one direction without sacrificing generality. 
The circuit generates high by activating switch $\mathrm{S_{out-s}}$, and low by activating switch $\mathrm{S_{out-p}}$. 
The diode prevents reverse current from flowing through the body diode of $\mathrm{S_{out-s}}$.

\subsubsection{Confirmation of autonomous maintenance of capacitor voltage}
\label{Confirmation of autonomous maintenance of capacitor voltage}
We demonstrate the transmission of power packets and the modifications in voltages of each capacitor installed in part $\alpha$--$\gamma$.

\begin{figure}[t]
  \centering
  \includegraphics[width = 0.6\hsize]{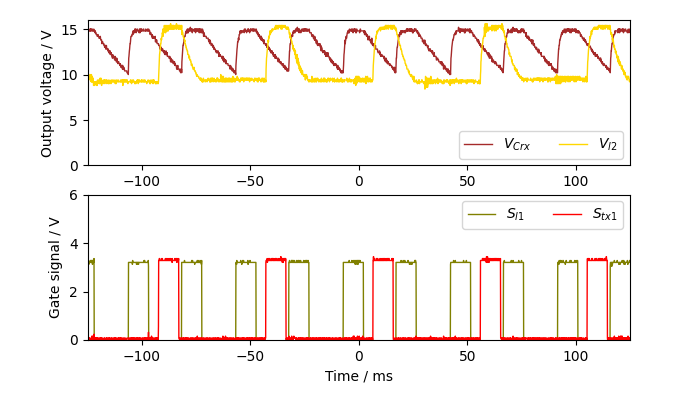}
  \caption{State of switches and voltage of capacitors in part $\alpha$.}
  \label{fig9}
\end{figure}
Figure~\ref{fig9} depicts the voltages $V_\mathrm{l1}$ and $V_\mathrm{tx}$ of the capacitors $C_\mathrm{l1}$ and $C_\mathrm{tx}$ in part $\alpha$ and the gate signal of the switches $S_\mathrm{l1}$ and $S_\mathrm{tx1}$ that controlled the route of the power packets. 
It is observed that $V_\mathrm{l1}$ and $V_\mathrm{tx}$ were sustained above the threshold voltages. 
The voltages $V_\mathrm{l1}$ and $V_\mathrm{tx}$ elevated when switches $S_\mathrm{l1}$ and $S_\mathrm{tx1}$ were driven. 
This demonstrates that capacitors $C_\mathrm{l1}$ and $C_\mathrm{tx}$ effectively received power packets and were charged. 
Furthermore, switching operation of $S_\mathrm{l1}$ and $S_\mathrm{tx1}$ did not overlap at any time. 
This result correlates to the setup that the transmission of power packets to $C_\mathrm{l1}$ is prioritized (see Section~\ref{Operation flow in connected systems} for the details).

\begin{figure}[t]
  \centering
  \includegraphics[width = 0.6\hsize]{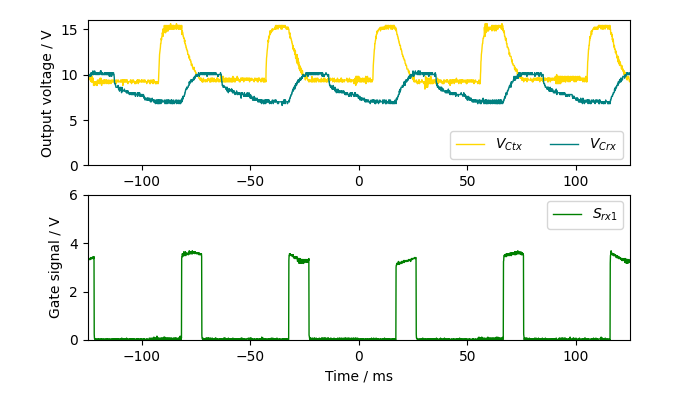}
  \caption{State of switches and voltage of capacitors in part $\beta$.}
  \label{fig10}
\end{figure}
Figure~\ref{fig10} depicts the voltages $V_\mathrm{tx}$ and $V_\mathrm{rx}$ of the capacitors $C_\mathrm{tx}$ and $C_\mathrm{rx}$ in part $\beta$ and the gate signal of the switch $S_\mathrm{rx}$ that controlled the power packet reception of the router rx.  
Comparing the top and bottom graphs shows that $V_\mathrm{tx}$ declined and $V_\mathrm{rx}$ elevated while $S_\mathrm{rx}$ was on. 
This implies that power packets were wirelessly transmitted successfully from router tx to router rx. 
It can also be validated that $S_\mathrm{rx}$ turned off when $V_\mathrm{tx}$ attained the threshold voltage. 
This implies that the system satisfied the constraints defined in Section~\ref{Operation flow in connected systems}, which hampers the output of power packets under the threshold voltage.

\begin{figure}[t]
  \centering
  \includegraphics[width = 0.6\hsize]{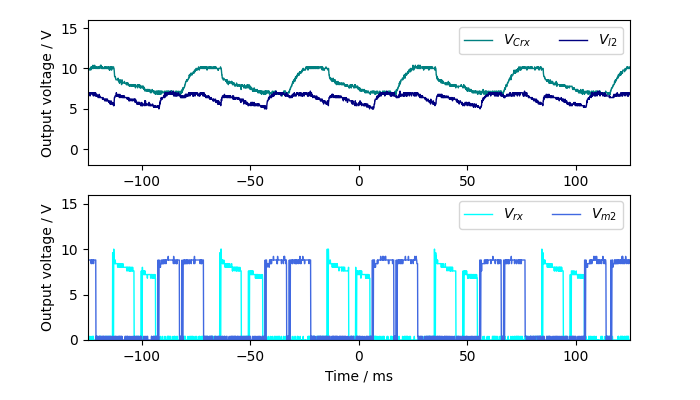}
  \caption{Power packets and voltage of capacitors in part $\gamma$.}
  \label{fig11}
\end{figure}
Figure~\ref{fig11} depicts the voltages $V_\mathrm{rx}$ and $V_\mathrm{l2}$ of the capacitors $C_\mathrm{rx}$ and $C_\mathrm{l2}$ in part $\gamma$ and voltage waveforms of power packets outputted from routers rx and m2. 
When $V_\mathrm{l2}$ dropped below the threshold voltage, router rx transferred power packets to router m2 so that $V_\mathrm{l2}$ was kept above the threshold. 
Now let us concentrate on the operation around $t=$\SI{25}{ms} when $V_\mathrm{rx}$ attains the threshold voltage.
Router rx stopped outputting the power packets, and simultaneously, router m2 started sending power packets. 
These findings suggest that the selective routing protocol specified in Section~\ref{Operation flow in connected systems} worked; the load sent the demand signal to the router rx at first, and if no packet was transmitted, then sent to the router m2. 

In \fref{fig11}, there exists a possible difference between the voltage of the power packet and $V_\mathrm{l2}$.
This was induced by the forward voltage drop across the diode installed to prevent backflow current.
This loss can be repressed by using a switch instead of a diode.
Thus, there is no impact on the verification of the principle.

\begin{figure}[t]
  \centering
  \includegraphics[width = 0.6\hsize]{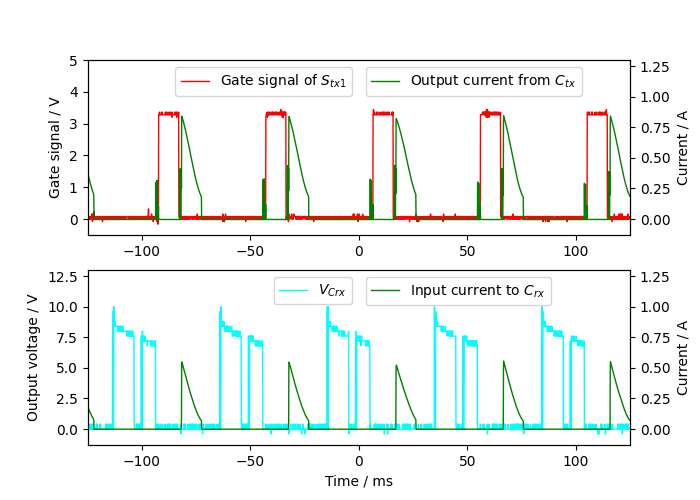}
  \caption{Input / output current and voltage of $C_\mathrm{tx}$ and $C_\mathrm{rx}$.}
  \label{fig12}
\end{figure}
Figure~\ref{fig12} depicts the gate signal of $S_\mathrm{tx1}$, output current from $C_\mathrm{tx}$, input current to $C_\mathrm{rx}$, and the output voltage waveforms of router rx. 
When $C_\mathrm{tx}$ was outputting current, $C_\mathrm{rx}$ was receiving current. 
This implies that the transmitted power packet was received without failure.
Since router tx did not output power packets when $S_\mathrm{tx1}$ was on, $S_\mathrm{tx1}$ and $S_\mathrm{tx2}$ were driven solely. 
Similarly, $S_\mathrm{rx1}$ and $S_\mathrm{rx2}$ were driven solely. 
From the above findings, it can be deduced that the connected system achieves the load voltage maintenance with wireless power supply between local systems 1 and 2 by following the control procedure defined in Section~\ref{Operation flow in connected systems}.

\subsubsection{Association between the percentage of power supply and the utilized power source}
\label{Association between the amount of power supply and the utilized power source}
The percentage of power transferred on the wireless channel depends on the distance between the transceiver/receiver coils. 
Hence, in the previous experiment's setup, increasing the coil gap reduced the power supply capability from the local system 1 to 2. 
The proposed control scheme of the routers can accommodate such a gap change by choosing an appropriate supply channel.  
To test this operation, we compare the amount of wireless power transmission and the power source selection in the local system 2 at various distances between the coils. 
We set three cases with different distances: (i) \SI{50}{mm} (the same as in the previous experiment), (ii) \SI{100}{mm}, and (iii) $>$ \SI{250}{mm}.
The setup in case iii is supposed to be large enough to prevent wireless transmission. 

\begin{figure}[t]
  \centering
  \includegraphics[width = 0.6\hsize]{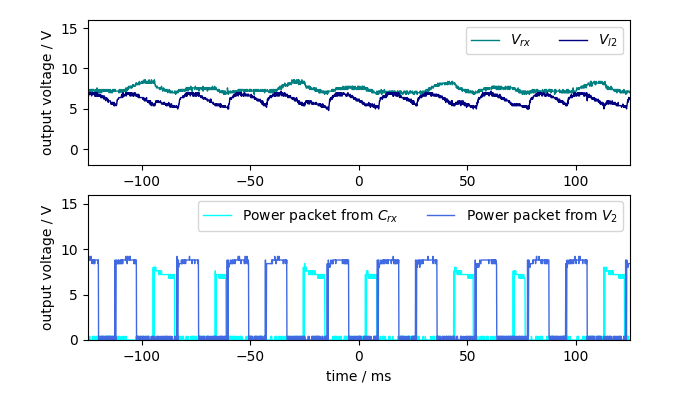}
  \caption{Power packets and voltage of capacitors in part $\gamma$ of case (ii) : gap \SI{100}{mm}.}
  \label{fig13}
\end{figure}
\begin{figure}[t]
  \centering
  \includegraphics[width = 0.6\hsize]{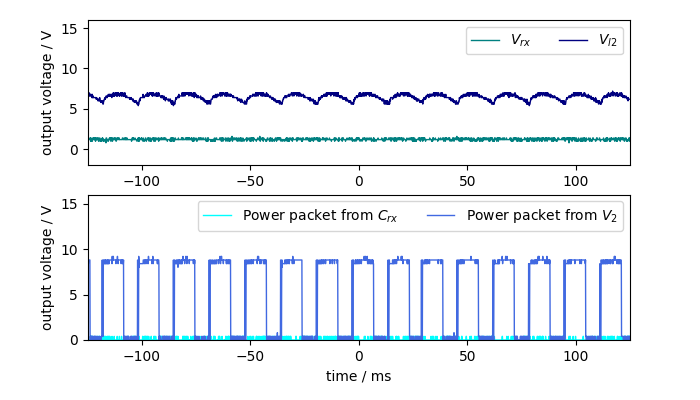}
  \caption{Power packets and voltage of capacitors in part $\gamma$ of case (ii) : gap \SI{250}{mm}.}
  \label{fig14}
\end{figure}
Figures~\ref{fig13} and \ref{fig14} depict the voltage $V_\mathrm{rx}$ and $V_\mathrm{l2}$ and the power packets output by routers rx and m2 in cases ii and iii. 
Please refer to \fref{fig11} for the result in case i. 
The larger the distance between the coils, the less frequently the router rx outputted power packets and the lower its average voltage got. 
On the other hand, $V_\mathrm{l2}$ maintained above the threshold in all cases. 

\begin{table}[t]
  \caption{Input/output power of the routers in local system 2 at each gap.}
  \label{inoutsub2}
  \begin{center}
  \begin{tabular}{cccccc}
  \hline
  \multirow{2}{*}{Case} & \multirow{2}{*}{Gap} & Router rx & Router rx & Router m2 &  Total \\
   & & input & output & output & output\\ \hline
  i & \SI{50}{mm} &  \SI{0.50}{W} &   \SI{0.46}{W} &   \SI{0.73}{W} & \SI{1.19}{W}\\ 
  ii & \SI{100}{mm} &   \SI{0.20}{W} &  \SI{0.17}{W} &   \SI{0.94}{W} & \SI{1.11}{W}\\ 
  iii & $>$ \SI{250}{mm} &   \SI{0.00}{W} &  \SI{0.00}{W} &   \SI{1.13}{W} & \SI{1.13}{W} \\ 
  \hline
  \end{tabular}
  \end{center}
\end{table}
\tref{inoutsub2} demonstrates the average of the input/output power of router rx and the output power of router m2 during the measured time \SI{250}{ms} for different distances. 
The input/output power of router rx fell and the output power of router m2 rose as the distance became larger. 
Meanwhile, the total output power of router rx and router m2 had a slight change. 
This finding implies that the output power of router m2 compensates for the fall in the output power of router rx. 

From the above, it is asserted that the load voltage can be sustained autonomously by the proposed distributed control scheme.
Even when the amount of wireless transmission falls, the local system compensated for it with a wired supply.

\section{Conclusion}
\label{Conclusion}
In this paper, we developed a platform for wireless power packet transmission for power management among numerous local systems. 

First, we proposed a novel power packet router configuration capable of wireless transmission. 
The ASK modulating circuit is installed on the router's output side for both information and power transmission, with the power packet serving as a power source. 
The input side includes a demodulation circuit for both information and power receipt. 
The circuit shifts between a signal demodulation circuit and a power rectifier circuit to read the header and receive the payload power, respectively.
Not only does the switching configuration separate the incoming signal and power, but it also reduces unnecessary power consumption during the receiving operation. 

Using this router, we then verified the wireless power packet routing following the information tag. 
Physical tag attachment and wireless power packet time-division multiplexing allowed receiving routers to distinguish the power packet based on its destination address. 
The result shows that the proposed configuration allows for the selective transmission of wireless power packets between multiple nearby local systems. 
This prevents interference with the irrelevant power supply. 

Next, we considered flexible coordination of inter- and intrasystem power management. 
The former was accomplished through the wireless transmission of power packets, while the latter was accomplished through a wired supply. 
For this purpose, we created a distributed control scheme for the routers.
A local system transmitted power packets wirelessly to another when it had enough power while keeping the voltage of its load as a top priority. 
The experiments revealed that the two types of operation were coordinated successfully. 
Furthermore, the proposed distributed control scheme chose an appropriate supply channel based on the power interaction availability between the local systems. 
We validated this operation by altering the gap between the coils of the two local systems, demonstrating that the inter- or intrasystem power management was successfully chosen to satisfy the local loads' demand. 

From the above verifications, we deduce that wireless power packet transmission can improve power management capability in a connected power packet dispatching system by selectively cooperating wired and wireless power packet transmission.

\section*{Acknowledgments}
This work was partially supported by JSPS KAKENHI 20H02151, JST-OPERA Program no. JPMJOP1841, and SIP Cross Ministerial Strategic Innovation Promotion Program no.18088028.


\end{document}